\documentclass{pasj00}
\SetRunningHead{T.Ohnishi et al.}{X-Ray Spectrum of G\,359.1$-$0.5}
\begin{document}
\title{X-Ray Spectrum of a Peculiar Supernova Remnant G\,359.1$-$0.5}
\author{Takao \textsc{Ohnishi},\altaffilmark{1}
  Katsuji \textsc{Koyama},\altaffilmark{1}
  Takeshi Go \textsc{Tsuru},\altaffilmark{1}
  Kuniaki \textsc{Masai},\altaffilmark{2}
  Hiroya \textsc{Yamaguchi},\altaffilmark{3}
  and
  Midori \textsc{Ozawa}\altaffilmark{1}
}
\altaffiltext{1}{Department of Physics, Kyoto University, Kitashirakawa-oiwake-cho, Sakyo-ku, Kyoto 606--8502}
\email{ohnishi@cr.scphys.kyoto-u.ac.jp}
\altaffiltext{2}{Department of Physics, Tokyo Metropolitan University, 1--1 Minami-Osawa, Hachioji, Tokyo 192--0397}
\altaffiltext{3}{RIKEN(The Institute of Physical and Chemical Research), 2--1 Hirosawa, Wako, Saitama 351--0198}


\maketitle
\begin{abstract}
We present the Suzaku results of a supernova remnant (SNR), G\,359.1$-$0.5 in the direction of the Galactic center region.
From the SNR, we find prominent K-shell lines of highly ionized Si and S ions, together with unusual structures at 2.5--3.0 and 3.1--3.6~keV.
No canonical SNR plasma model, in either ionization equilibrium or under-ionization, can explain the structures.
The energies and shapes of the structures are similar to those of the radiative transitions of free electrons to the K-shell of He-like Si and S ions (radiative recombination continuum: RRC).
The presence of the strong RRC structures indicates that the plasma is in over-ionization.
In fact, the observed spectrum is well fitted with an over-ionized plasma model.
The best-fit electron temperature of 0.29~keV is far smaller than the ionization temperature of 0.77~keV, which means that G\,359.1$-$0.5 is in extreme condition of over-ionization.
We report some cautions on the physical parameters, and comment possible origins for the over-ionized plasma.
\end{abstract}

\section{Introduction}
G\,359.1$-$0.5 was discovered by \citet{Downes_1979} as a shell-like object with the 4.875~GHz radio band, and was suggested to be a supernova remnant (SNR).
\citet{Reich_1984} observed G\,359.1$-$0.5 with the 2.695~GHz and 4.75~GHz radio bands, and found a complete shell structure.
Together with the spectral index ($\alpha=-0.37$) and the liner polarization, they confirmed this object to be an SNR.
\citet{Uchida_1992a} found a $^{12}$CO molecular ring that is concentric with the shell of G\,359.1$-$0.5.
The molecular ring has high radial velocities between $-$60 and $-$190~km~s$^{-1}$.
Furthermore, \citet{Uchida_1992b} found a ring structure of G\,359.1$-$0.5 in the 1.42~GHz band, surrounded with the H\emissiontype{I} emission ring having large negative velocities of $-$75 to $-$190~km~s$^{-1}$.
These large velocity differences of about 100~km~s$^{-1}$ would be due to the shear motions in the Galactic center (GC) region, and hence G\,359.1$-$0.5 is likely located in the GC region.

The non-thermal radio filament, named the "Snake" (e.g. \cite{Gray_1991}; \cite{Gray_1995}), is located at the projected position of G\,359.1$-$0.5.
Along the shell of G\,359.1$-$0.5 and perpendicular to the Snake, \citet{Lazendic_2002} discovered shocked molecular hydrogen (H$_2$).
The intriguing fact is that an OH (1720~MHz) maser source, a signature of an SNR/molecular cloud interaction, is located in the H$_2$ cloud at the projected intersection of G\,359.1$-$0.5 with the Snake \citep{Yusef-Zadeh_1995}.
Thus G\,359.1$-$0.5 and the Snake may be physically interacting, although it could be an artifact of superposition of the disparate components along the line of sight.
These radio morphologies place G\,359.1$-$0.5 as one of the enigmatic SNR possibly related to the extreme environment of the GC.

The X-ray emission inside the radio shell was first discovered with ROSAT \citep{Egger_1998}.
The ASCA observation also found the center-filled X-rays and claimed prominent K-shell emission lines from He-like ions of Si and those from H-like ions (Ly$\alpha$) of S \citep{Bamba_2000}.
From the radio and X-ray morphologies, G\,359.1$-$0.5 is likely to be a mixed-morphology SNR \citep{Rho_1998}, although the X-ray image data were poor for this concrete statement.

The line features of ASCA were very peculiar, because S should be ionized far higher than Si, although S and Si are synthesized in the same layer of a progenitor star.
In fact, the X-ray spectrum required a two-component collisional ionization equilibrium (CIE) plasma of 0.6 and 4.4~keV temperatures and an extreme over-abundance of S ($>$ 38~solar) in the higher temperature plasma.
Thus the X-ray spectrum of G\,359.1$-$0.5 is also enigmatic.
These peculiar features in the radio morphologies and in the X-ray spectrum may be related with each other and may require a new scenario for the structure and evolution of the SNR.

We observed G\,359.1$-$0.5 with Suzaku \citep{Mitsuda_2007}.  
With the excellent energy resolution, the large effective area, and the low background for diffuse emission, the X-ray Imaging Spectrometer (XIS; \cite{Koyama_2007b}) provide the first opportunity to entangle the peculiarity of G\,359.1$-$0.5.
In this paper, we report the high quality X-ray spectra and detailed analyses, and discuss the nature of G\,359.1$-$0.5.
All of the errors are at the 1$\sigma$ confidence level throughout this paper, unless otherwise mentioned.

\section{Observations and Data Reduction}
The G\,359.1$-$0.5 region was observed as a part of the Galactic Center Survey Project of Suzaku on 2008 September 14--16.
We also used its vicinity to obtain a background spectrum and to examine surroundings of the G\,359.1$-$0.5 \citep{Bamba_2009}.
The observation logs are given in table \ref{observationID}.

We used the XIS data for the imaging and spectral analyses.
Details on the XIS are found in \citet{Koyama_2007b}.
Further information of XIS on the calibration, qualification, and so on, is found in separate papers (e.g., \cite{Uchiyama_2009}).
Although the XIS consists of four CCD cameras with three front-illuminated (FI) CCDs (XIS\,0, XIS\,2, and XIS\,3) and one back-illuminated (BI) CCD (XIS\,1), we did not use one of the FI CCDs (XIS\,2) because of the loss of the proper function since 2006 November.

The data processing and screening were made using the processing version 2.2.11.22, {\tt xispi} software version 2008-06-06, and the calibration database version 2009-07-17.
After the processing and screening under the standard criteria\footnote{http://heasarc.nasa.gov/docs/suzaku/processing/criteria\_xis.html}, the exposure time were 57.7, 70.5, and 72.5~ks for the G\,359.1$-$0.5 and two vicinity fields, respectively.

\begin{table*}
  \begin{center}
    \caption{Observation logs.}
    \label{observationID}
    \begin{tabular}{ccccc}
      \hline
      Observation ID & Start date & R.A. & Dec. & Exposure$^*$ \\
      & & \multicolumn{2}{c}{(J2000.0)} & (ks) \\
      \hline
      503012010 & 2008-09-14 & $\timeform{266D.2992}$ & $\timeform{-29D.9408}$ & 57.7 \\
      502016010 & 2008-03-02 & $\timeform{266D.2249}$ & $\timeform{-30D.1096}$ & 70.5 \\
      502017010 & 2008-03-06 & $\timeform{266D.4705}$ & $\timeform{-30D.0867}$ & 72.5 \\
      \hline
      \multicolumn{5}{@{}l@{}}{\hbox to 0pt{\parbox{106mm}{\footnotesize
	    \par\noindent
	    \footnotemark[$*$] Effective exposure of the screened XIS data.
	  }\hss}}
    \end{tabular}
  \end{center}
\end{table*}

\section{Analysis}
We use HEAsoft version 6.6.1 and SPEX \citep{Kaastra_1996} version 2.02.02 in the imaging and spectral analyses. 
The Non-X-ray Background (NXB) is made using {\tt xisnxbgen} \citep{Tawa_2008} and subtracted from the raw data.
XIS response and auxiliary files are created using {\tt xisrmfgen} and {\tt xissimarfgen} \citep{Ishisaki_2007}, respectively.
Since we use SPEX for the spectral fitting, we make format conversion using {\tt trafo} version 1.02.

\subsection{Image}
\begin{figure}
  \begin{center}
    \FigureFile(80mm,){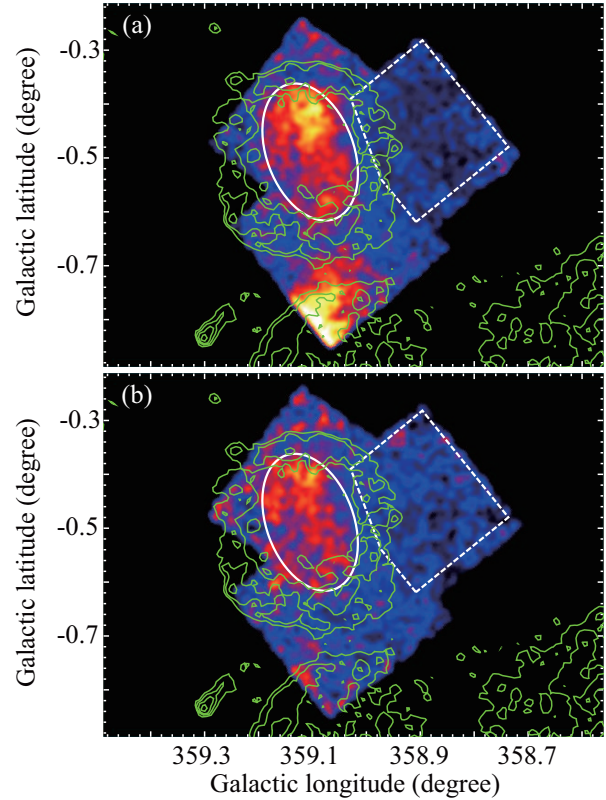}
  \end{center}
  \caption{XIS images in the (a) 1.5--2.3~keV and (b) 2.5--3.0~keV bands.
    The source and background regions are shown as a solid ellipse and a dotted pentagonal (white), respectively.
    The radio map by the VLA 90~cm observation is superimposed in green contours.}
  \label{image}
\end{figure}

Referring to the X-ray spectrum of G\,359.1$-$0.5 by \citet{Bamba_2000}, we show the X-ray images in the (a) 1.5--2.3~keV and (b) 2.5--3.0~keV bands.
After the NXB subtraction and the subsequent corrections of the exposures and the vignetting effects, all the XIS images are merged to increase statistics.
We shift the Suzaku coordinates referring to the XMM-Newton sources \citep{Watson_2009}, because the Suzaku nominal position error is $\lesssim\timeform{20''}$ \citep{Uchiyama_2008}, on the other hand, that of the XMM-Newton catalog is $\sim\timeform{2''}$.
The name of the reference stars detected with XMM-Newton and Suzkau, their positions ($l$, $b$) and differences ($\Delta l$, $\Delta b$) are listed in table \ref{reference}.
After these fine-shifts of the Suzaku coordinates, we make the images in figures \ref{image}a and \ref{image}b.
The figures are binned with $\timeform{8''.3}\times\timeform{8''.3}$ and smoothed with a Gaussian kernel of $\sigma=\timeform{0.'70}$.
The radio map by the VLA 90~cm observation \citep{LaRosa_2000} is superimposed with green contours.

\begin{table*}
  \begin{center}
    \caption{Reference sources used in the position adjustments of the Suzaku data.}
    \label{reference}
    \begin{tabular}{cccccccc}
      \hline
      Observation ID & Reference source & \multicolumn{2}{c}{Source position} & \multicolumn{2}{c}{Source position} & \multicolumn{2}{c}{Difference} \\
      \multicolumn{1}{c}{Suzaku} & \multicolumn{1}{c}{XMM-Newton} & \multicolumn{2}{c}{XMM-Newton} & \multicolumn{2}{c}{Suzaku} & & \\
      & & $l$ & $b$ & $l$ & $b$ & $\Delta l$ & $\Delta b$ \\
      \hline
      503012010 & J174539.4$-$300139 & $\timeform{359D.0726}$ & $\timeform{-0D.5756}$ & $\timeform{359D.0733}$ & $\timeform{-0D.5765}$ & $\timeform{0D.0007}$ & $\timeform{-0D.0008}$ \\
      502016010 & J174458.3$-$300655 & $\timeform{358D.9205}$ & $\timeform{-0D.4951}$ & $\timeform{358D.9207}$ & $\timeform{-0D.4952}$ & $\timeform{0D.0002}$ & $\timeform{-0D.0001}$ \\
      502017010 & J174558.0$-$295738 & $\timeform{359D.1648}$ & $\timeform{-0D.5982}$ & $\timeform{359D.1674}$ & $\timeform{-0D.6003}$ & $\timeform{0D.0026}$ & $\timeform{-0D.0021}$ \\
      \hline
    \end{tabular}
  \end{center}
\end{table*}

In figures \ref{image}a and \ref{image}b, we find diffuse X-rays inside the radio shell of G\,359.1$-$0.5 (green contours).
No excess X-ray relative to the surrounding region is found on the radio shell.
This is the first convincing evidence that G\,359.1$-$0.5 is a mixed-morphology SNR.
We designate the ``source region'' by the solid ellipse centered on ($l$, $b$)=($\timeform{+359D.10}$, $\timeform{-0D.49}$) with the semimajor and semiminor axes of $\timeform{7'.9}$ and $\timeform{4'.9}$, respectively.
The bright emission in the lower corner of figure \ref{image}a is a part of another SNR G\,359.0$-$0.9 (\cite{Bamba_2000}; \cite{Bamba_2009}).
Since the present observations cover only a small portion of the SNR, we do not treat this SNR in this paper.

\subsection{Spectra}
The X-ray spectra of G\,359.1$-$0.5 are made from the source region.
Since this SNR is located in the direction of the GC, the Galactic Center Diffuse X-ray (GCDX; e.g., \cite{Koyama_2007a}) is a major background for the spectrum.
We therefore select the dotted pentagonal region as shown in figures \ref{image}a and \ref{image}b, where no X-ray excess is found (hereafter, the background region).
The mean Galactic latitude is almost the same as that of the source region, which may compensate the strong flux dependence of GCDX on the Galactic latitude.
The source and background spectra are made by subtracting the NXB.
After the correction of the differences of the effective areas and the effective exposures including the vignetting effects for both the source and background region, we subtract the background spectra from the source spectra.
To improve the photon statistics, the background-subtracted source spectra of the two FIs (XIS\,0 and XIS\,3) are merged because the response functions are almost identical.
The background-subtracted FI spectrum of G\,359.1$-$0.5 (the source region) is shown in figure \ref{CIE}a.

We see no significant X-rays in the hard X-ray band above 4~keV, and hence fit the FI and BI spectra in the 1.0--4.0~keV band simultaneously, although we show only the FI spectrum in the figures for brevity. 
 
\subsubsection{Collisional Ionization Equilibrium plasma}
\begin{figure}
  \begin{center}
    \FigureFile(80mm,){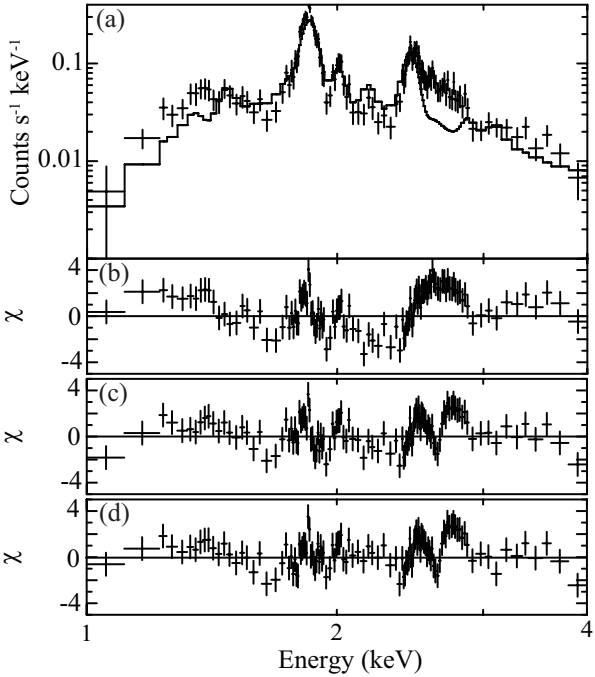}
  \end{center}
  \caption{(a) Background-subtracted FI spectrum of G\,359.1$-$0.5 with the best-fit 1-CIE plasma model.
    (b) Residuals from the best-fit 1-CIE plasma model.
    (c) Residuals from the best-fit 2-CIE model with the common abundances.
    (d) Same as (c), but the Si and S abundances are unlinked between the two components.
}
  \label{CIE}
\end{figure}

Using the code of SPEX, we fit the spectra with the model of an optically thin thermal plasma in CIE affected by photoelectric absorption with the cross-sections of \citet{Morrison_1983} (1-CIE fit).
The abundances of Si and S ($Z_{\rm Si}$ and $Z_{\rm S}$) are free parameters, but those of the other elements are fixed at the solar values \citep{Lodders_2009}.
The best-fit 1-CIE model is given in figure \ref{CIE}a with the solid line, while the data residuals are shown in figure \ref{CIE}b.
Residuals are found around the energies of 2.5--3.0 and 3.1--3.6~keV, although the latter is marginal.
Therefore, this model is rejected with the large $\chi^2$/degrees of freedom (d.o.f.) of 585/191.

We then fit the spectra with a two-component CIE model with the common absorption for both the CIE components (2-CIE fit).
The abundances of Si and S are free parameters but are linked between the two components.
The abundances of the other elements are fixed at the solar values.
The data residuals from the best-fit 2-CIE model are shown in figure \ref{CIE}c.
The residuals around 2.5--3.0~keV are still found even by this model with the $\chi^2$/d.o.f. of 365/189.
We thus try the 2-CIE model fitting but with unlinked Si and S abundances.
This is the same model accepted for the ASCA spectrum \citep{Bamba_2000}.
We, however, find the unacceptable fitting with $\chi^2$/d.o.f. of 360/187, remaining the significant residuals as is shown in figure \ref{CIE}d.

The best-fit values of the cooler and hotter plasma temperatures are 0.20~$^{+0.02}_{-0.01}$ and 1.5~$\pm0.1$~keV, respectively, which are inconsistent with those reported by \citet{Bamba_2000}.
This inconsistency would be due to the different sensitivity in the energy band below Mg-Ly$\alpha$ energy ($<$~1.5~keV) between the Suzaku XIS and ASCA GIS (Gas Imaging Spectrometers; \cite{Ohashi_1996}).
Although \citet{Bamba_2000} fitted the spectra in the energy band of 1--1.5~keV, the sensitivity of this energy band is very small.
The XIS, on the other hand, has higher sensitivity in the same energy band.
We therefore fit the XIS spectra with the best-fit model and values reported by \citet{Bamba_2000}, and find large residuals at the low energy and around 2.5--3.0~keV with $\chi^2$/d.o.f. of 821/193.

\subsubsection{Under-ionized plasma}
Since almost all the plasma of young SNRs are in under-ionization, we fit the spectra of G\,359.1$-$0.5 with one or two-component under-ionized plasma model affected by photoelectric absorption using the SPEX code.
The abundances of Si and S are free parameters, while the abundances of the other elements are fixed at the solar values.
These model are rejected with the large $\chi^2$/d.o.f. of 585/190 and 365/188.
In the both cases, the best-fit ionization temperatures ($kT_{\rm z}$) are the same as those of the electron temperature ($kT_{\rm e}$).
These fittings are essentially the same as those of the CIE model.
We therefore do not proceed further analysis on the under-ionized plasma.

\subsubsection{Over-ionized plasma}
\begin{figure}
  \begin{center}
    \FigureFile(80mm,){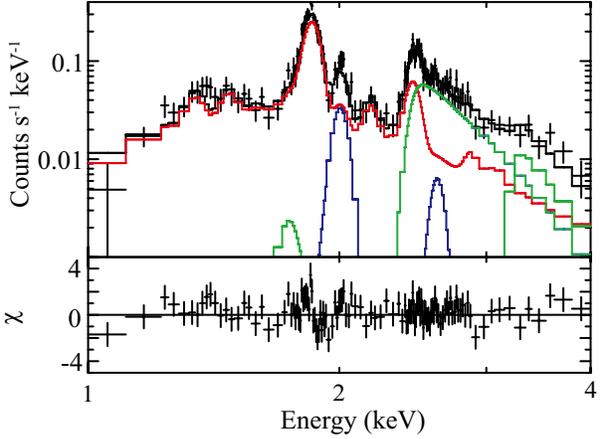}
  \end{center}
  \caption{Same spectrum as figure \ref{CIE}a, but fitted with Model~A.
    The CIE plasma model, the RRC, and the Ly$\alpha$ lines of the Si and S are shown as red, green, and blue lines, respectively.
  }
  \label{RRC}
\end{figure}

The energies of the residuals found in the 1-CIE fit correspond to the K-shell binding potentials of the He-like Si (2439~eV) and He-like S (3225~eV).
These structures are similar to the radiative transitions of free electrons to the K-shell of ions (radiative recombination continuum: RRC) found in IC\,443 and W49B (\cite{Yamaguchi_2009}; \cite{Ozawa_2009}).
From the energies of the structures, these are thought to be the RRC for the transitions of free electrons to the K-shell of He-like Si and S ions.
We therefore examine if the RRC are real or not, by adding the RRC of He-like ions of Si and S to a 1-CIE plasma model.
For consistency, we also add the Ly$\alpha$ lines of Si and S with the fixed energy of the theoretical values at 2006 and 2622~eV, respectively (hereafter, Model~A).
The electron temperature values for the RRC of Si and S are common free parameters.
Then the $\chi^2$/d.o.f. value is dramatically improved to 224/186, where the best-fit spectrum of Model~A are shown in figure \ref{RRC}.
The best-fit CIE temperature and abundances are $kT_{\rm CIE}$ of 0.62~$^{+0.11}_{-0.06}$~keV, $Z_{\rm Si}$ of 4.1~$^{+0.6}_{-0.5}$~solar and $Z_{\rm S}$ of 4.2~$^{+1.3}_{-1.1}$~solar, respectively, while the best-fit RRC temperature is $kT_{\rm e}$ of 0.31~$^{+0.05}_{-0.04}$~keV.

As is shown in figure \ref{RRC}, we find the RRC of He-like Si ions mainly contribute to the structure, of which the ASCA observation claimed the Ly$\alpha$ lines of S ions \citep{Bamba_2000}.
Unlike IC\,443 and W49B, the fluxes of the RRC in G\,359.1$-$0.5 are nearly one order of magnitude higher than the underlying bremsstrahlung (see Model~A in figure \ref{RRC}), and $kT_{\rm e}$ determined from the width of the RRC structures is significantly away from $kT_{\rm CIE}$ determined mainly from the bremsstrahlung continuum in the 1-CIE plasma.
Therefore, such a simple analysis cannot give physical quantities properly, such as total energy and metal abundances.

\begin{figure}
  \begin{center}
    \FigureFile(80mm,){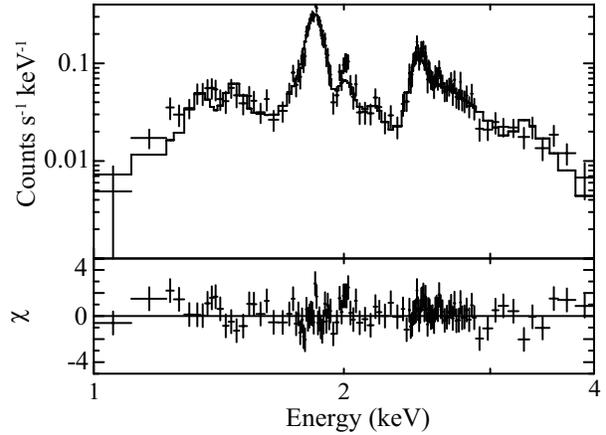}
  \end{center}
  \caption{Same spectrum as figure \ref{CIE}a, but fitted with Model~B.
  }
  \label{overionized}
\end{figure}

We apply an over-ionized plasma model\footnote{For this analysis, we used a code named "CIE" in the SPEX package.
Although the name of "CIE" is misleading, this code can treat the electron ($kT_{\rm e}$) and ionization ($kT_{\rm z}$) temperatures as independent free parameters, and hence can describe plasma states in over-ionization ($kT_{\rm z}>kT_{\rm e}$).} for the first time to the whole SNR spectrum, in which electron and ionization temperatures are treated as independent free parameters (hereafter, Model~B).
The abundances of Mg, Si, and S are free parameters, while the other elements are fixed at the solar values.
Model~B gives the best-fit $\chi^2$/d.o.f. of 223/189, which is marginally acceptable (null hypothesis probability $=$ 0.05).
The best-fit model and parameters of Model~B are shown in figure \ref{overionized} and table \ref{overionized_table}, respectively.

\begin{table}
  \begin{center}
    \caption{The best-fit parameters of Model~B.}
    \label{overionized_table}
    \begin{tabular}{lll}
      \hline
      \multicolumn{1}{c}{Component} & \multicolumn{1}{c}{Parameter} & \multicolumn{1}{c}{Value} \\
      \hline
      Absorption & & \\
      & $N_{\rm H}$ $\left(10^{22}~{\rm H~cm}^{-2}\right)$ & 2.0 $^{+0.3}_{-0.2}$ \\
      Over-ionized plasma & & \\
      & $kT_{\rm e}$ $\left({\rm keV}\right)$ & 0.29 $\pm0.02$ \\[2pt]
      & $kT_{\rm z}$ $\left({\rm keV}\right)$ & 0.77 $^{+0.09}_{-0.08}$ \\[2pt]
      & $Z_{\rm Mg}$ $\left({\rm solar}\right)$ & 3.4 $^{+1.2}_{-0.9}$ \\[2pt]
      & $Z_{\rm Si}$ $\left({\rm solar}\right)$ & 12 $^{+3}_{-2}$ \\[2pt]
      & $Z_{\rm S}$ $\left({\rm solar}\right)$ & 17 $^{+5}_{-4}$ \\[2pt]
      & VEM$^*$ & 7.2 $^{+2.3}_{-1.8}$ \\
     \hline
      \multicolumn{2}{@{}l@{}}{\hbox to 0pt{\parbox{85mm}{\footnotesize
	    \par\noindent
	    \footnotemark[$*$] Volume emission measure.
            The unit is ${\int}n_{\rm e}n_{\rm p}dV/4{\pi}D^2$ $\left(10^{11}~{\rm cm}^{-5}\right)$, where $n_{\rm e}$, $n_{\rm p}$, $V$, and $D$ are the electron and proton densities (cm$^{-3}$), the emitting volume (cm$^3$), and the distance to the source (cm), respectively.
          }\hss}}
    \end{tabular}
  \end{center}
\end{table}

\subsubsection{Spatial structure of the spectra}
\label{spatial}
Although the size of G\,359.1$-$0.5 is reasonably large compared to the angular resolution of the XIS, the poor photon statistics limit the study of the spatial structure of the over-ionized plasma.
We therefore simply divide the source region of G\,359.1$-$0.5 into two at $b=\timeform{-0D.5}$: the near side of the Galactic plane (here, near side) and the far side from the plane (far side), and examine the spectral difference.
We extract the spectra from the near side and the far side of the source region separately, and fit each spectrum with the same Model~B of the fixed column density ($N_{\rm H}$) and the abundances of Mg, Si, and S ($Z_{\rm Mg}$, $Z_{\rm Si}$, and $Z_{\rm S}$) at the best-fit value of Model~B (table \ref{overionized_table}).

The best-fit parameters of the far side are $kT_{\rm e}$ of 0.26 $\pm0.03$~keV, $kT_{\rm z}$ of 0.80 $^{+0.14}_{-0.12}$~keV, and volume emission measure (VEM) of 3.0 $\pm0.3$~$\times10^{11}~{\rm cm}^{-5}$.
On the other hand, those of the near side are $kT_{\rm e}$ of 0.31 $\pm0.02$~keV, $kT_{\rm z}$ of 0.75 $^{+0.07}_{-0.06}$~keV, and VEM of 4.7 $\pm0.4$~$\times10^{11}~{\rm cm}^{-5}$.

Although within the statistical errors, we see hints that the electron temperature in the far side is smaller than that in the near side and vice-versa for the ionization temperature.
For the quantitative comparison, we define the ionization parameter (IP) as the ratio of ionization temperature to the electron temperature ($kT_{\rm z}/kT_{\rm e}$), then we see that the IP in the far side of 3.0 $^{+0.4}_{-0.3}$ is slightly higher than that in the near side of 2.4 $\pm0.1$.
From the best-fit VEM, we confirm that the density in the far side is smaller than that in the near side.

\begin{figure}
  \begin{center}
    \FigureFile(80mm,){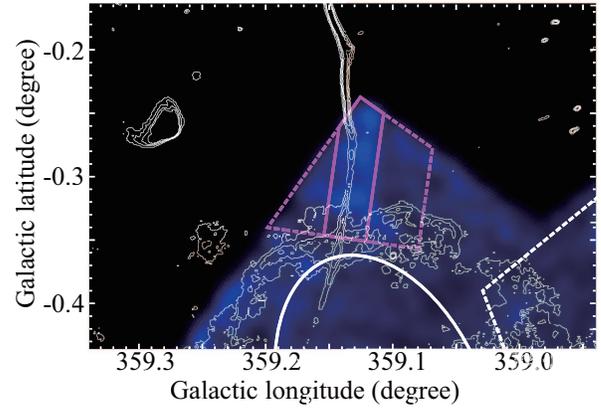}
  \end{center}
  \caption{XIS close-up image in the 4.0--8.0~keV band.
    To avoid contamination by the emission from the calibration source, we excluded the 5.7--6.7~keV band.
    The radio map by the VLA 20~cm observation is superimposed in the contours \citep{Yusef-Zadeh_2004}.
  }
  \label{snake}
\end{figure}

Since the near side may be interacting with the Snake (figure \ref{snake}), the non-thermal radio filament, the IP of this region might be influenced by high energy electrons, which may emit non-thermal X-rays by synchrotron radiation. 
Although we find no hint of excess X-rays in the hard X-ray 4.0--8.0~keV band as shown in figure \ref{snake}, we try to estimate the upper-limit of the hard X-ray flux.

We make the X-ray spectrum from the pentagonal region (magenta), where the background is taken from the adjacent region shown by the dotted magenta lines in figure \ref{snake}.
Since the statistics is limited, we fit the spectrum with a power-law model of fixed index 2.0 affected by photoelectric absorption of the fixed $N_{\rm H}=2.0~\times~10^{22}~{\rm H~cm}^{-2}$, the best-fit parameter of Model~B.
Then we constrain the flux upper-limit to be 2$\times10^{-14}$~ergs~cm$^{-2}$~s$^{-1}$ at the 90\% confidence level.

Using this upper-limit of the X-ray flux, we estimate the energy of the non-thermal electrons.
Since the spectral index and the flux derived from 1446 and 4790~MHz vary along the length of the Snake, we adopt the spectral index to be 0, based on figure 12 of \citet{Gray_1995} within latitude from $-15\timeform'$ to $-21\timeform'$, which corresponds to the pentagonal region.
Assuming that the size of the Snake in the radio observations within the pentagonal region is $\timeform{6'}$ $\times$ $\timeform{0.2'}$, we estimate the flux density to be 70~mJy at 1446~MHz.
We adopt the magnetic field strength of 88~$\mu$G, which is the minimum magnetic field of the Snake estimated by the radio observations \citep{Gray_1995}.
Then the best-fit srcut model, which represents the synchrotron spectrum from an exponentially cut-off power-law distribution of electrons (\cite{Reynolds_1998}; \cite{Reynolds_1999}), gives the upper-limit of the cut-off electron energy to be 1~GeV.

\section{Discussion}
We find that the peculiar X-ray spectrum of G\,359.1$-$0.5 is nicely reproduced by the over-ionized plasma model with $kT_{\rm e}$ of 0.29 and $kT_{\rm z}$ of 0.77~keV unlike the previous X-ray study \citep{Bamba_2000}, in which the X-ray spectra required the two-component CIE plasma.
Due to the limited energy resolution and photon statistics of ASCA, the RRC of Si was misunderstood as the Ly$\alpha$ line of S, and thus the non-existent high-temperature (4.4~keV) S-rich plasma would be claimed.
In the following sections, we discuss possible implications based on this over-ionized plasma model.

\subsection{Physical parameters in the over-ionized model}
\label{physical}
We note that the Si and S abundances obtained from the Model~A are different from those of Model~B.
This would demonstrate the limitations of simple analysis of Model~A in high IP cases.

We must also pay attention to the result of Model~B that continuum X-rays due to the RRC from heavy elements are significant compared to the bremsstrahlung.
Since the RRC emission in 1.0--2.0~keV is mainly due to the abundant elements, O and Ne, the bremsstrahlung flux from the H and He atoms is greatly affected by the O and Ne abundances.
We therefore cannot properly estimate the metal abundances relative to H.

In order to estimate the abundance errors due to putative large RRC of O and Ne, we fit the spectra with Model~B, but the abundances of O and Ne are fixed as 4~solar, near the same value of Mg.
The apparent abundances of Mg, Si, and S are 9.9, 34, and 48~solar ($\chi^2$/d.o.f. of 227/189), which are largely different from table \ref{overionized_table} (for 1-solar O and Ne).
On the other hand if we estimate the abundances of Mg, Si, and S relative to O or Ne, then the relative abundances are modified to 2.5, 8.6, and 12~solar, for Mg, Si, and S, respectively.
These values are similar to those of Model~B (table \ref{overionized_table}).

With the same reason, the VEM calculated from the bremsstrahlung of H and He is 2.4$\times10^{11}~{\rm cm}^{-5}$ in 4-solar O and Ne abundances, which is about a third of the Model~B.
On the other hand, $N_{\rm H}$, $kT_{\rm e}$, and $kT_{\rm z}$ are estimated to be 2.0~$\times10^{22}~{\rm H~cm}^{-2}$, 0.29, and 0.79~keV, respectively, thus the uncertainty of these parameters are small.

The best-fit column density of 2.0$\times10^{22}$ H cm$^{-2}$ (table \ref{overionized_table}) is equal to or slightly smaller than that of other X-ray sources near to the GC with the same Galactic coordinates of this SNR \citep{Sakano_1999}.
Hence, this SNR would be located in or front of the GC.
This is consistent with the previous radio observations.
Hereafter, we assume that the SNR is located in the GC region with the distance of 8.5~kpc.
The shape of the source region is approximated with an ellipse of 12.0$\times$19.6~pc$^2$ (see figure \ref{image}).
Assuming that the total kinetic energy of ions (protons) in the plasma is equal to $kT_{\rm e}$, the sound velocity is estimated to be 3$\times10^{7}$~cm~s$^{-1}$.
Dividing the major axis by the sound velocity, we estimate the dynamical time scale to be 7$\times10^4$~yr.

\subsection{Origin of the over-ionized plasma}
The most remarkable discovery is the extremely over-ionized plasma from the mixed-morphology SNR G\,359.1$-$0.5.
The ionization temperature of $kT_{\rm z}=0.77$~keV is similar to, but electron temperature of $kT_{\rm e}=0.29$~keV is far lower than those of the middle-aged SNR.
Although the number of samples for over-ionized SNRs are very small, G\,359.1$-$0.5 is unusual compared to the other mixed morphology SNRs, IC\,443 and W49B.
The possible scenarios for the origin of over-ionized plasmas in IC\,443 and W49B have been discussed (\cite{Kawasaki_2002}; \cite{Yamaguchi_2009}; \cite{Miceli_2006}; \cite{Miceli_2010}).
Unfortunately, the observational results of G\,359.1$-$0.5 are very limited due to the poor statistics for the quantitative discussion on the origin, we hence briefly comment and address some plausible scenarios.

{\it Thermal conduction}: \citet{Kawasaki_2002} proposed that thermal conduction from a central hot plasma to a surrounding cool plasma caused the strong cooling of the electrons in IC\,443.
G\,359.1$-$0.5 was reported to be surrounded by the $^{12}$CO molecular ring and the H$\emissiontype{I}$ emission ring (\cite{Uchida_1992a}; Uchida et al. 1992b).
Furthermore, OH maser (1720~MHz) sources, tracers of shock activity, are located at the projected position of the SNR shell (\cite{Yusef-Zadeh_1995}; \cite{Yusef-Zadeh_1996}).
Since the emissions of the surrounding $^{12}$CO, H$\emissiontype{I}$, and OH masers are more intense at the shell of the near side than that of the far side, this part of G\,359.1$-$0.5 would interact strongly with the molecular cloud.
Then the thermal conduction from the plasma to the surrounding molecular cloud might cause the cooling of the electrons more effectively and produce a more over-ionized plasma in the near side.
Thus, the IP in the near side may be higher than that in the far side.

{\it Adiabatic cooling}: \citet{Yamaguchi_2009} proposed that the drastic adiabatic expansion of the shocked gas causes the rapid cooling of electrons \citep{Itoh_1989}.
The supernova explodes in a dense circumstellar medium, and then the gas is shock-heated and significantly ionized at the initial phase of the SNR evolution.
The subsequent outbreak of the blast wave to a low-density interstellar medium (ISM) causes the drastic adiabatic expansion of the shocked gas and the resultant rapid cooling of the electrons.
This scenario predicts that the IP is higher in the region with lower electron density (far side), since the adiabatic expansion in the low-density ISM produces an over-ionized plasma.

{\it Suprathermal electrons}: Over-ionized plasmas were found in big solar flares with hard X-ray tails (e.g. \cite{Tanaka_1986}).
\citet{Kato_1992} proposed that a few percent of the suprathermal electrons, which produce the hard X-rays by bremsstrahlung, caused effective ionization.
Suprathermal electrons may be supplied by the radio filament Snake.
Since the most effective ionizations are made near a few 10~keV (by electrons) and several 10~MeV (by protons), our constraint of the upper-limit of electron energy of 1~GeV does not rule-out this possibility. 

We find a hint that the IP is larger in the far side than in the near side.
This fact may favor the second scenario, the adiabatic cooling of electrons.
For quantitative calculation on the origin of this extremely over-ionized plasma, detailed observations of the spatial structure is encouraged.

\bigskip
The authors thank Hideki Uchiyama for his useful comments on X-ray absorption and Masayoshi Nobukawa, Makoto Sawada, and Shinya Nakashima for improving the draft. 
This work was supported by the Grant-in-Aid for the Global COE Program "The Next Generation of Physics, Spun from Universality and Emergence" from the Ministry of Education, Culture, Sports, Science and Technology (MEXT) of Japan. 
This work was also supported by Scientific Research B No. 20340043 (TT), Scientific Research C No. 22540253 (KM), and Challenging Exploratory Research No. 2054019 (KK), from Japan Society for the Promotion of Science(JSPS).

\end{document}